\documentclass[twocolumn,aps,epsfig,nofootinbib]{revtex4}

\usepackage{graphicx}
\usepackage{epstopdf}
\usepackage{latexsym}
\usepackage{amssymb}
\usepackage{amsmath}
\usepackage{color}
\usepackage{mathrsfs}
\usepackage[center]{subfigure}

\begin{document}

\newcommand{\bq}{\begin{equation}}
\newcommand{\eq}{\end{equation}}
\newcommand{\bqn}{\begin{eqnarray}}
\newcommand{\eqn}{\end{eqnarray}}
\newcommand{\nb}{\nonumber}
\newcommand{\lb}{\label}
\newcommand{\PRL}{Phys. Rev. Lett.}
\newcommand{\PL}{Phys. Lett.}
\newcommand{\PR}{Phys. Rev.}
\newcommand{\PRD}{Phys. Rev. D}
\newcommand{\CQG}{Class. Quantum Grav.}
\newcommand{\JCAP}{J. Cosmol. Astropart. Phys.}
\newcommand{\JHEP}{J. High. Energy. Phys.}
\newcommand{\PLB}{Phys. Lett. B}

\title{Kerr Geodesics Following the Axis of Symmetry\footnote{This article is dedicated to the memory of 
our longtime friend and collaborator, Professor William Bowen Bonnor, who  passed away on August 17,  2015.}}

\author{J. Gariel$^1$}
\email{jerome.gariel@obspm.fr}

\author{ N. O. Santos $^{1,2}$}
\email{n.o.santos@qmul.ac.uk} 

\author{Anzhong Wang $^{3, 4, 5}$}
\email{anzhong$\_$wang@baylor.edu}

\affiliation{
$^{1}$Sorbonne Universit\'es, UPMC Universit\'e Paris 06, LERMA, UMRS8112 du CNRS,
Observatoire de Paris-Meudon 5, Place Jules Janssen, F-92195 Meudon Cedex, France\\
$^{2}$School of Mathematical Sciences, Queen Mary, University of London, London E1 4NS, UK\\
$^{3}$GCAP-CASPER, Department of Physics, Baylor University, Waco, Texas 76798-7316, USA\\
$^{4}$ Institute  for Advanced Physics $\&$ Mathematics, Zhejiang University of Technology, 
Hangzhou 310032,  China\\
$^{5}$ Departamento de F\'{\i}sica Te\'orica, Instituto de F\'{\i}sica, UERJ, 20550-900, Rio de Janeiro, Brazil}

\begin{abstract}
{ We present here   the general expressions for the acceleration of massive test particles along  the symmetry axis of the Kerr metric, and then study the main properties of this acceleration in different regions of the spacetime. In particular, we show that there exists a region near the black hole in which  the gravitational field is repulsive.  We provide possible physical  interpretations about the role of this effect in terms of the different conserved parameters. The studies of these geodesics are important not only to understand better the structure of the Kerr spacetime but also to its use as a possible mechanism for the production of extragalactic jets. Our results are obtained with the help of expressing the geodesics of the Kerr spacetime in terms of the Weyl coordinates. }  
\end{abstract}

\maketitle

\section{Introduction}

In this paper,  we study timelike geodesics of the Kerr space-time  that follow its axis of symmetry. These geodesics revealed great importance to build a possible mechanism to explain the production of very high energy particles and of extragalactic jets stemming from the centre of galaxies \cite{Bicak,Williams,Williams1,Felice,Gariel,Pacheco,Gariel1,Gariel2}. It is widely accepted that the centre of galaxies encloses a Kerr black hole and particles, from the accretion disk, can fall into it by crossing the ergosphere. Inside the ergosphere they may undergo a Penrose process producing new particles that can be ejected with high energies \cite{Williams,Williams1,Schnittman}. These particles can follow collimated geodesics near the axis of symmetry of the black hole, and thus { form } the   extragalactic jets.  In particular, it was showed   the possibility of an almost perfect collimation of geodesics along the axis of symmetry of the Kerr spacetime by using Weyl coordinates and producing specific asymptotic jet radii \cite{Gariel}, and among these geodesics there exist observable particles with very high energies (theoretically infinite) \cite{Pacheco}.

In \cite{Gariel1} this jet production mechanism has been used to compare its predictions concerning the profile of the jet to recent observed data from the galaxy Messier 87, and found in good agreement with the observations \cite{Asada}. It has been also shown in \cite{Gariel1}, among its results, that by assuming that the angular momentum per unit length $a$ takes the value that maximizes the ergospheric volume, there is a remarkable compatibility with the recent data obtained from Messier 87 \cite{Li}.
In \cite{Gariel2} it was demonstrated  that the Penrose process can produce with great efficiency particles with very high energies.  

In order to have a better visibility of the geodesics in the Kerr space-time,  specially of the ones moving along the symmetry  axis, in this paper we study them by transforming the geodesic equations usually presented in the spheroidal Boyer-Lindquist coordinates into  the cylindrical Weyl  coordinates. In this system of coordinates, it becomes very clear  that the intriguing repulsive effect, already noted in cylindrical systems for both static and stationary spacetimes \cite{Herrera,Opher}, is also present in stationary axial symmetric spacetimes. 
The advantage of the cylindrical Weyl  coordinates  in such studies was already  taken  to highlight a quasi-perfect collimation, i.e., parallel to the symmetry axis, for a class of geodesics  in the Kerr space-time \cite{Gariel}. In particular,  it was argued that astrophysical jet collimation may arise from the geometry of rotating black holes and the presence of high-energy particles resulting from a Penrose process, without the help of magnetic fields. This is very attractive and provides a special mechanism for formation of jets with the participation of neutral particles. 

%
%

  In this paper, we shall focus our attention on particle acceleration along the symmetry axis near the ergosphere, and  show explicitly  how the potential energy is converted into kinetic energy that allows the escape of energetic particles. The rest of the paper is organized in the following way. In section 2 we briefly review  the Kerr geodesics in the Weyl coordinates, while in section 3 we study the acceleration of particles along the axis $z$.  This acceleration on the ergosphere surface is further investigated in section 4. Particles moving along the axis $z$ and attaining large $z$ distances are investigated  in sections 5 and 6. The paper finishes with a brief conclusion in section 7.

\section{Geodesics in Kerr Spacetime}
In this section, to initiate our studies we shall provide a brief review over   the main results of the timelike geodesics of the Kerr space-time  in both  of the  Boyer-Lindquist   \cite{Chandra} and Weyl systems of coordinates \cite{Gariel}. For more details, we refer readers to \cite{Chandra,Gariel}.

The Kerr metric given in the Boyer-Lindquist spheroidal coordinates $\bar r$, $\theta$ and $\phi$ reads
\begin{eqnarray}
ds^2=-\left(\frac{{\bar r}^2-2M{\bar r}+a^2}{{\bar r}^2+a^2\cos^2\theta}\right)(dt
-a\sin^2\theta\, d\phi)^2 \nonumber\\
+\frac{\sin^2\theta}{{\bar r}^2+a^2\cos\theta^2}
\left[a\, dt-({\bar r}^2+a^2)\,d\phi\right]^2 \nonumber\\
+({\bar r}^2+a^2\cos^2\theta)\left(\frac{d{\bar r}^2}{{\bar r}^2-2M{\bar r}+a^2}+d\theta^2\right),
\label{1}
\end{eqnarray}
where $M$ and $aM$ are, respectively, the mass and the angular momentum of the source. In this paper, we shall choose units such that $c=G=1$,
 where $G$ is Newton's constant of gravitation. Rescaling the $\bar r$ coordinate as $r={\bar r}/M$, two of the timelike geodesics equations take the forms, 
\begin{eqnarray}
&& {\dot r}^2= \frac{a_4r^4+a_3r^3+a_2r^2+a_1r+a_0}{M^2\left[r^2+\left(\frac{a}{M}\right)^2\cos^2\theta\right]^{2}},
\label{2}\\
&& {\dot\theta}^2=\frac{b_4\cos^4\theta+b_2\cos^2\theta+b_0}{M^2\left(1-\cos^2\theta\right)\left[r^2
+\left(\frac{a}{M}\right)^2\cos^2\theta\right]^{2}}, \label{3}
\end{eqnarray}
where the dot stands for differentiation with respect to an affine parameter, and  the coefficients $a_i$'s and $b_j$'s are defined by
\begin{eqnarray}
a_0=-\frac{a^2Q}{M^4}=-\left(\frac{a}{M}\right)^2b_0, \label{4}\\
a_1=\frac{2}{M^2}\left[(aE-L_z)^2+Q\right], \label{5}\\
a_2=\frac{1}{M^2}\left[a^2(E^2-1)-L_z^2-Q\right], \label{6}\\
a_3=2, \label{7}\\
a_4=E^2-1, \label{8}
\end{eqnarray}
and
\begin{eqnarray}
b_0=\frac{Q}{M^2}, \label{9}\\
b_2=\frac{1}{M^2}\left[a^2(E^2-1)-L^2_z-Q\right]=a_2, \label{10}\\
b_4=-\left(\frac{a}{M}\right)^2(E^2-1)=-\left(\frac{a}{M}\right)^2a_4, \label{11}
\end{eqnarray}
where $E$, $L_z$ and $Q$ are integration constants. 
The other two geodesics equations are
\begin{widetext}
\begin{eqnarray}
&& M{\dot\phi}=\left\{2\frac{a}{M}Er+\left[r^2-2r+\left(\frac{a}{M}\right)^2\cos^2\theta\right]
\frac{L_z}{M\sin^2\theta}\right\}  
\left[r^2-2r+\left(\frac{a}{M}\right)^2\right]^{-1}
\left[r^2+\left(\frac{a}{M}\right)^2\cos^2\theta\right]^{-1}, \label{b12} \\
&& {\dot t}=\left[\left\{\left[r^2+\left(\frac{a}{M}\right)^2\right]^2
-\left(\frac{a}{M}\right)^2\left[r^2-2r+\left(\frac{a}{M}\right)^2\right]\sin^2\theta\right\}E  
-2\frac{aL_z}{M^2}r\right]\nb\\
&&  ~~~~~~ \times \left[r^2-2r+\left(\frac{a}{M}\right)^2\right]^{-1}
\left[r^2+\left(\frac{a}{M}\right)^2\cos^2\theta\right]^{-1}.
\label{b13}
\end{eqnarray}
\end{widetext}
 Here Chandrasekhar's $\delta_1$ has been set equal to one (See  p. 327 of \cite{Chandra}), as in this paper we consider only the  timelike geodesics. Assuming that the affine parameter is the proper time $\tau$ along the geodesics, then these equations imply that  a unit mass for the test particle is assumed, so that $E$ and $L_z$ have the usual significance of total energy and angular momentum about the $z$-axis, and $Q$ is the corresponding Carter constant.  
  In this paper we consider only particles on unbound geodesics with $E\geq 1$  \cite{Chandra}.

In order to understand better the physical meaning of the Kerr geodesics, as mentioned previously, in the following  we shall use the Weyl cylindrical coordinates $\rho$, $z$ and $\phi$, which are more revealing and, in a way, quite natural for  axially symmetric spacetimes. The dimensionless Weyl cylindrical coordinates, in multiples of geometrical units of mass $M$, are given by  \cite{Gariel}
\begin{equation}
\rho=\left[(r-1)^2-A\right]^{1/2}\sin\theta, \;\; z=(r-1)\cos\theta, \label{12}
\end{equation}
where
\begin{equation}
A=1-\left(\frac{a}{M}\right)^2. \label{13}
\end{equation}
Inversely,  we have  
\begin{eqnarray}
r=\alpha +1, \label{14}\\
\sin\theta=\frac{\rho}{(\alpha^2-A)^{1/2}}, \;\; \cos\theta=\frac{z}{\alpha},\\
\label{15}
z=\left(1-\frac{\rho^2}{\alpha^2-A}\right)^{1/2}\alpha, \label{15a}
\end{eqnarray}
with
\begin{equation}
\alpha=\frac{1}{2}\left\{\left[\rho^2+(z+\sqrt{A})^2\right]^{1/2}+
\left[\rho^2+(z-\sqrt{A})^2\right]^{1/2}\right\}. \label{16}
\end{equation}
Here we have assumed $A\geq 0$, and taken the root of the second degree equation obtained from (\ref{15}) for the function $\alpha(\rho,z)$ that allows the extreme black hole limit $A=0$. The other root in this limit is $\alpha=0$.

From (\ref{16}) it can be shown that the curves of constant $\alpha$ (constant $r$) are ellipses with semi-major axis $\alpha$ and eccentricity $e={\sqrt A}/\alpha$ in the $(\rho,z)$ plane, which implies that for large $\alpha$ these approach  to circles. Note that $\rho=0$ consists of the rotation axis $\theta=0$ or $\pi$ together with the ergosphere surface.

Now, with the help of Eqs.(\ref{14}) and (\ref{15}) we can rewrite the geodesics equations (\ref{2}) and (\ref{3}) in terms of $\rho$ and $z$ coordinates as
\begin{widetext}
\begin{eqnarray}
&& M{\dot\rho}=\frac{1}{U}\left[\frac{P\alpha^3\rho}{\alpha^2-A}
+\frac{S(\alpha^2-A)z}{\alpha\rho}\right], \label{17}\\
&& M{\dot z}=\frac{1}{U}\,(Pz-S)\alpha, \label{18}\\
&& M{\dot\phi}=\left\{2\frac{a}{M}E(\alpha+1)+[U-2\alpha^2(\alpha +1)]\frac{L_z}{M(\alpha^2-z^2)}\right\}\frac{\alpha^2}{U(\alpha^2-A)}, \label{c18}\\
&& {\dot t}=\left\{E\left[(\alpha+1)^2+\left(\frac{a}{M}\right)^2\right]^2-2\frac{aL_z}{M^2}(\alpha+1)\right\}\frac{\alpha^2}{U(\alpha^2-A)}  
-\left(\frac{a}{M}\right)^2\frac{E}{U}(\alpha^2-z^2), \label{c19}
\end{eqnarray}
\end{widetext}
where
\begin{eqnarray}
&& P=\left[a_4(\alpha+1)^4+a_3(\alpha+1)^3+a_2(\alpha+1)^2\right.\nb\\
&& ~~~~~~~~ \left. +a_1(\alpha+1)+a_0\right]^{1/2}, \label{19}\\
&& S=-(b_4z^4+b_2\alpha^2z^2+b_0\alpha^4)^{1/2}, \label{20}\\
&& U=(\alpha+1)^2\alpha^2+\left(\frac{a}{M}\right)^2z^2, \label{21}
\end{eqnarray}
with the sign being chosen to indicate outgoing particle geodesics. 
Instead of the $\rho$ geodesics,  we can express it in terms of $\alpha$. Using (\ref{16}), (\ref{17}) and (\ref{18}) we find
\begin{equation}
M{\dot\alpha}=\frac{P\alpha^2}{U}. \label{22}
\end{equation}

\section{Acceleration along the symmetry axis}

In order to study the acceleration of a test massive particle along the $z$ axis,   taking the second proper time derivative of (\ref{18}), we find
\begin{widetext}
\begin{eqnarray}
MU^2{\ddot z}=U({\dot P}z-{\dot S})\alpha
+\left[UP-2\left(\frac{a}{M}\right)^2(Pz-S)z\right]\alpha{\dot z} 
+(Pz-S)\left[U-2\alpha^2(\alpha+1)(2\alpha +1)\right]{\dot \alpha}. \label{23}
\end{eqnarray}
From (\ref{7}), (\ref{19}) and (\ref{20}) we have
\begin{eqnarray}
MU({\dot P}z-{\dot S})=\frac{1}{2}\left[4a_4(\alpha +1)^3+3a_3(\alpha +1)^2
+2a_2(\alpha +1)+a_1\right]\alpha^2z
-2PS\alpha+(2b_4z^2+b_2\alpha^2)\alpha z. \label{24}
\end{eqnarray}
Substituting (\ref{18}), (\ref{22}) and (\ref{24}) into (\ref{23}) we obtain
\begin{eqnarray}
 \frac{M^2U^3}{\alpha^2}\;{\ddot z}&=& \frac{U}{2}\left[4a_4(\alpha +1)^3+3a_3(\alpha+1)^2
+2a_2(\alpha +1)+a_1\right]\alpha z
+U\left[(2b_4z^2+b_2\alpha^2)z-2PS\right] \nonumber\\
&& +2(Pz-S)\left[UP-\left(\frac{a}{M}\right)^2(Pz-S)z\right]
-2(Pz-S)P\alpha^2(\alpha+1)(2\alpha+1), \label{25}
\end{eqnarray}
or
\begin{eqnarray}
\frac{M^2U^3}{\alpha^2}\;{\ddot z}&=&\frac{U}{2}\left[4a_4(\alpha+1)^3+3a_3(\alpha+1)^2
+2a_2(\alpha+1)+a_1\right]\alpha z
\nonumber\\
&& +U(2b_4z^2+b_2\alpha^2)z
-2P^2\alpha^3(\alpha+1)z-2\left(\frac{a}{M}\right)^2S^2z-2PS\alpha^2(\alpha+1). \label{26}
\end{eqnarray}
Substituting (\ref{19}), (\ref{20}) and (\ref{21}) into (\ref{25}) we obtain
\begin{eqnarray}
\frac{M^2U^3}{\alpha^2}\;{\ddot z}&=& 
-\frac{1}{2}\left[a_3(\alpha+1)^3+2a_2(\alpha+1)^2+3a_1(\alpha+1)+4a_0\right]
\alpha^3(\alpha+1)z \nonumber\\
&& +\frac{1}{2}\left(\frac{a}{M}\right)^2\left[4a_4(\alpha+1)^3+3a_3(\alpha+1)^2+2a_2(\alpha+1)
+a_1\right]\alpha z^3 \nonumber\\
&& +(2b_4z^2+ b_2\alpha^2)\alpha^2(\alpha+1)^2z
-\left(\frac{a}{M}\right)^2(b_2z^2+2b_0\alpha^2)\alpha^2z  -2PS\alpha^2(\alpha+1). \label{27}
\end{eqnarray}
Now,  considering (\ref{4}), (\ref{7}), (\ref{10}) and (\ref{11}) and substituting into (\ref{27}) we have
\begin{eqnarray}
\frac{M^2U^3}{\alpha^3}\;{\ddot z}
&=& -\left[(\alpha+1)^4+\left(a_2+\frac{3}{2}\;a_1\right)(\alpha+1)^2+2a_0\right]
\alpha^2z 
+\left(\frac{a}{M}\right)^2\left[(2a_4+3)(\alpha+1)^2+a_2+\frac{1}{2}a_1\right] z^3\nb\\
&& -2PS\alpha(\alpha+1). \label{28}
\end{eqnarray}
\end{widetext}

Writing (\ref{20}) with the help of (\ref{15a}) we obtain
\bqn
S&=&-\left[b_0+b_2+b_4-(b_2+2b_4)\frac{\rho^2}{\alpha^2-A}\right.\nb\\
&& \left. 
~~~~ +b_4\frac{\rho^4}{(\alpha^2-A)^2}\right]^{1/2}\alpha^2, \label{30}
\eqn
where, with (\ref{9}), (\ref{10}) and (\ref{11}), we have
\begin{eqnarray}
b_0+b_2+b_4=-\left(\frac{L_z}{M}\right)^2\leq 0, \label{31}\\
b_2+2b_4=-\frac{1}{M^2}\left[a^2(E^2-1)+L_z^2+Q\right]. \label{32}
\end{eqnarray}

\section{Test particles being ejected from ergosphere surface}

The ergosphere has outer surface given in the Boyer-Lindquist coordinates  (\ref{1})
 by 
\begin{equation}
r=1+\left[1-\left(\frac{a}{M}\right)^2\cos^2\theta\right]^{1/2}. \label{32a}
\end{equation}
Transforming it into the Weyl coordinates (\ref{14}-\ref{15a}), we find that it can be cast in the form, 
\begin{equation}
z^2=\left[1-\rho_e^2\left(1-\frac{\rho}{\rho_e}\right)\right]\left(1-\frac{\rho}{\rho_e}\right), \label{33a}
\end{equation}
where $\rho_e=a/M$ is the value of $\rho$ for which the ergosphere cuts the equatorial plane $z=0$, and $\rho=(0,\rho_e)$.

Now we calculate the acceleration along the $z$ axis given by Eq.(\ref{28}) for particles leaving the ergosphere in the two limits $\rho=0$ and $\rho=\rho_e$.
For $\rho=0$ we have from (\ref{16}) and (\ref{32a})
\begin{equation}
\alpha=z=\sqrt{A}, \label{34a}
\end{equation}
producing from (\ref{20})
\begin{equation}
S=-(b_4+b_2+b_0)^{1/2}A, \label{35a}
\end{equation}
and by (\ref{31}) we see that only geodesics with $L_z=0$ can exist leaving the ergosphere at $\rho=0$. From (\ref{28}) and (\ref{34a}) we have
\begin{equation}
2(\sqrt{A}+1)^2M^2\ddot{z}=-\sqrt{A}\left[\sqrt{A}+1+\rho_e^2(E^2-1)+\frac{Q}{M^2}\right]. \label{36a}
\end{equation}
From (\ref{36a}) we have that for $\rho=0$ on the ergosphere surface the field along $z$ is always attractive if the Carter constant is positive, $Q>0$, or, being negative, $Q<0$, if
\begin{equation}
\frac{|Q|}{M^2}<\sqrt{A}+1+\rho_e^2(E^2-1). \label{37a}
\end{equation}
In the extreme Kerr black hole, $\rho_e=1$ or $A=0$, we have from (\ref{36a}) ${\ddot z}=0$.

For $\rho=\rho_e$ on the ergosphere outer surface (\ref{33a}), we have
\begin{equation}
z=0, \;\; \alpha=1, \label{38a}
\end{equation}
and from (\ref{20}) we find
\begin{equation}
S=-\frac{\sqrt Q}{M}. \label{39a}
\end{equation}
From (\ref{39a}) we see that in order to exist geodesics leaving (\ref{38a}) we must have $Q>0$ and from (\ref{28}) we obtain
\bqn
{\ddot z}&=& \left[16M^2+4\rho_e^2(2E^2-1)-8\rho_eE\frac{L_z}{M}-\rho^2_e\frac{Q}{M^2}\right]^{1/2}\nb\\
&& ~~ \times \frac{\sqrt Q}{16M^3}, \label{40a}
\eqn
with the condition
\begin{equation}
8\rho_eE\frac{L_z}{M}+\rho^2_e\frac{Q}{M^2}<16E^2+4\rho^2_e(2E^2-1). \label{41a}
\end{equation}
Hence, the field given by (\ref{40a}) is repulsive where the Carter constant plays a fundamental role. On one hand it increases the repulsive field ${\ddot z}>0$, while on the other hand the term $\rho^2_eQ/M^2$ decreases its repulsive character. To find out which value of $Q/M^2$ shall produce the highest repulsive field with respect to given $\rho_e$, $E$ and $L_z/M$, we calculate the extremum of (\ref{40a})
\bqn
 \frac{d(16M^2{\ddot z})^2}{d(Q/M^2)}&=& 16M^2+4\rho_e^2(2E^2-1)-8\rho_eE\frac{L_z}{M}\nb\\
&&  -2\rho^2_e\frac{Q}{M^2}=0. \label{42a}
\eqn
When $Q/M^2$ satisfies the above equation, it can be shown that such a value of $Q/M^2$ will produce a maximum of $\ddot z$, which is given by
\begin{equation}
{\ddot z}_{max}=\frac{1}{4\sqrt 2M^2\rho_e}\left[4M^2+\rho_e^2(2E^2-1)-2\rho_eE\frac{L_z}{M}\right]^{3/2}. \label{43a}
\end{equation}
{ In the neighbourhood of $\rho=\rho_e$ we have $\rho=\rho_e-\varepsilon$, where $0<\varepsilon\ll 1$, and the condition to be located on the ergosphere is $z=(\varepsilon/\rho_e)^{1/2}$, as one can see from (\ref{33a}).  With the help of  (\ref{28}) we see that the first term in the right hand side is of the order $\sqrt\varepsilon$, and the second one is  of the order $\varepsilon^{3/2}$, which can be neglected. On the other hand, from (\ref{16}) we have $\alpha=1-\varepsilon\rho_e$, so the third term in the right hand side of (\ref{28}) has two sub-terms, one is  positive and of zeroth-order  of $\varepsilon$, while the other is  of order $\varepsilon$. Hence,  in the neighbourhood of $\rho=\rho_e$ on the ergosphere surface the field is still repulsive, $\ddot z>0$, along the symmetry $z$.}

Since along the outer ergosphere surface at $\rho=\rho_e$ { and in its neighbourhood we have ${\ddot z}>0$, while} on the axis we have ${\ddot z}<0$, then   ${\ddot z}$ must change its sign at least once on the outer ergosphere surface.

\section{Particles moving near the equatorial plane}

{ On the equatorial plane, $z=0$, the acceleration of a test particle along $z$  is 
\begin{equation}
\ddot z=-\frac{2PS\alpha^4(\alpha+1)}{M^2U^3}. \label{s1}
\end{equation}
Its velocity along $\rho$- and $z$-directions are given by  Eqs.(\ref{17}) and (\ref{18}), which can be cast in the forms, 
\begin{equation}
\dot\rho=\frac{P\alpha^3\rho}{MU(\alpha^2-A)}, \;\; \dot z=-\frac{S\alpha}{MU}. \label{s2}
\end{equation}
Substituting (\ref{s2}) into (\ref{s1}) we obtain
\begin{equation}
\ddot z=\frac{2\rho{\dot\rho}{\dot z}}{[(\rho^2+A)^{1/2}+1](\alpha^2+A)}. \label{s3}
\end{equation}
First, let us note  that (\ref{s3}) is also valid  in the neighbourhood of $z$-axis, since the terms including $z\neq 0$ are of order higher than those given in (\ref{s3}). From (\ref{s3}) we can see that, if the particle is approaching the source, $\dot\rho<0$, then $\ddot z<0$, which means that there is a force that diminishes its speed along $z$; while if the particle is distancing the source, $\dot\rho>0$, then $\ddot z>0$ and there is a force that increases its speed along $z$. This result is valid not only for the Kerr spacetime, but also for the Schwarzschild one ($a=0$). We notice that the presence of $a$ enhances   this effect.

A similar result has been obtained for the Levi-Civita  and  Lewis spacetimes \cite{Herrera}, which describe the gravitational fields produced by an infinite static or stationary line mass. In these two cases the stationarity did not intervene, which is not the case in (\ref{s3}). A further similar result has also been obtained for the van Stockum spacetime \cite{Opher}, describing an infinite rotating dust cylinder.}

Lots  of works in axial symmetric spacetimes use the Weyl coordinates \cite{Exact}. 
Of course, in principle, any admissible coordinates, such as the Kerr-Schild cylindrical ones introduced in \cite{Bicak}, could be used, although it is well-known that the proper choice of coordinates for a specific problem can make the task significantly simplified.  In particular, the complicated differential nature of the transformations of the Boyer-Lindquist coordinates into the cylindrical Kerr-Schild  ones (See equations (29-33) in \cite{Bicak}) does not furnish an immediate link between these two  systems of coordinates. This in turn  prevents us from  finding the equivalence in the Kerr-Schild coordinates of our $\rho_1$ asymptotes (see (61) below). Nevertheless, the results obtained in \cite{Bicak} about the collimation and the repulsive character for high energy geodesics especially for naked singularities are consistent with  ours.

\section{Test particles attaining large $z$ distances}

In order to have geodesics attaining large distances along the axis, $z\gg\rho$,  from (\ref{30}), (\ref{31}) and (\ref{32}) we find that for $S$ real we must assume 
\begin{eqnarray}
L_z=0, \label{33}\\
-(b_2+2b_4)=\rho_e^2(E^2-1)+\frac{Q}{M^2}>0. \label{34}
\end{eqnarray}
Combining  (\ref{17}), (\ref{18}), (\ref{28}), (\ref{15a}) and (\ref{33}),  we obtain 
\begin{widetext}
\begin{eqnarray}
&& M{\dot\rho}=\frac{\alpha}{U}\left\{\frac{P\alpha^2\rho}{\alpha^2-A}-\left[b_4\rho^2
-(b_2+2b_4)(\alpha^2-A)\right]^{1/2}z\right\}, \label{36}\\
&& M{\dot z}=\frac{\alpha}{U}\left\{Pz+\left[b_4\frac{\rho^2}{\alpha^2-A}-(b_2-2b_4)\right]^{1/2}
\frac{\rho\alpha^2}{(\alpha^2-A)^{1/2}}\right\}, \label{33b}\\
&& \frac{M^2U^3}{\alpha^6}{\ddot z}=-\left[(\alpha+1)^4+\left(a_2+\frac{3}{2}a_1\right)(\alpha+1)^2+2a_0\right]
\left(1-\frac{\rho^2}{\alpha^2-A}\right)^{1/2} \nonumber\\
&&~~~~~~~~~~~~~~ +\rho_e^2\left[(2a_4+3)(\alpha+1)^2+a_2
+\frac{1}{2}\;a_1\right]\left(1-\frac{\rho^2}{\alpha^2-A}\right)^{3/2} \nonumber\\
&& ~~~~~~~~~~~~~~ +2P(\alpha+1)\left[b_4\frac{\rho^2}{\alpha^2-A}-(b_2+2b_4)\right]^{1/2}\frac{\rho}{(\alpha^2-A)^{1/2}}.
\label{35}
\end{eqnarray}
\end{widetext}
For  $z\gg\rho$ and $z\gg{\sqrt A}$, we find that  Eqs.(\ref{36})-(\ref{35}) become,
\begin{eqnarray}
M{\dot \rho}\approx-\left[\rho_e^2(E^2-1)+\frac{Q}{M^2}\right]^{1/2}\frac{1}{z} \label{35b},\\
M{\dot z}\approx (E^2-1)^{1/2}, \label{36b}\\
M^2{\ddot z}\approx -\frac{1}{z^2}, \label{36bb}
\end{eqnarray}
where (\ref{35b}) and (\ref{36b}) are identical to  (39) and (47) found  in \cite{Gariel}.  From (\ref{35b}) we find that
the speed of the particle in the $\rho$-direction decreases as $1/z$, while its speed $\dot z$ along the axial direction  also decreases and approaches  a constant.

\section{Test particles moving along the axis with $\rho=0$}

The geodesics along the $z$ axis with $\rho=0$ requires that $L_z=0$, as it can be seen  from (\ref{30}), while from (\ref{36}) we find that $b_2+2b_4=0$ or equivalently 
\begin{equation}
\rho_e^2(E^2-1)=-\frac{Q}{M^2},\label{37b}
\end{equation}
which in turn imposes $Q<0$, since here we have $E^2 >0$. This result corresponds to $\rho_1=0$, where
\begin{equation}
\rho_1=\rho_e\left[1+\frac{\mathcal Q}{a^2(E^2-1)}\right]^{1/2}, \label{37bb}
\end{equation}
for unbound geodesics with $L_z=0$  studied in \cite{Gariel}.
The acceleration given by (\ref{35}) along the symmetry axis $\rho=0$ then becomes
\begin{equation}
M^2{\ddot z}=-\left[(z+1)^2-\rho_e^2\right]\left[(z+1)^2
+\rho_e^2\right]^{-2}. \label{37}
\end{equation}
Since $(z+1)^2\geq\rho_e^2$ we have from (\ref{37}) that ${\ddot z}\leq 0$, which means that the field is attractive along the axis. However, it should be noted that
  $a$ diminishes the strength of its attractive character, as compared to the Newtonian limit, when $\rho_e=0$ and $Q=0$, producing
\begin{equation}
M^2{\ddot z}=-\frac{1}{(z+1)^2}. \label{38}
\end{equation}

\section{Conclusion}
We have calculated explicitly the acceleration $\ddot z$ along the axis of symmetry for the Kerr spacetime following its unbound geodesics and investigated the dependence of the  nature
of the acceleration  on the parameters $a$, $E$, $L_z$ and $Q$. In particular,  we have shown that at the limiting points of the ergosphere $(\rho=\rho_e,z=0)$ and $(\rho=0,z=\sqrt A)$ the signs of $\ddot z$ changes, showing that there must exist  at least one turning-point on the ergosphere outer surface, across which $\ddot z$ changes its signs. 
Repulsive forces near the symmetry  axis  have also been found in other  axisymmetric space-times \cite{Opher,Herrera}.   In the Kerr space-time, to produce such repulsive forces, we have found the special  role played by the spin $a$ (via the position $\rho_e$ of the ergosphere) and the Carter constant $Q$ (via the asymptotical direction $\rho_1$ of unbound geodesics).

{The presence of such a repulsive force along the axis and very near the ergosphere reinforces the idea that the early launching of jets could be initiated by the gravitational field of the rotating  black hole. This is strongly supported by  the recent observations of M87, which showed that the basis of the jet is very near to the ergosphere size \cite{Doeleman,Asada}.  Other evidences include   a quasi-perfect collimation \cite{Gariel} along $z$ with big energies \cite{Pacheco},  the possibility of a Penrose process \cite{Williams,Williams1,Schnittman}, and/or a BSW effect \cite{Banados}. Besides, this repulsive force can also be the source of cosmic rays of high energy, specially of neutrinos.

}

\section*{ Acknowledgements}

This work is supported in part by  Ci\^encia Sem Fronteiras, No. A045/2013 CAPES, Brazil (A.W.) and 
NSFC No. 11375153, China (A.W.).

\end{document}